\documentclass[
 reprint,
superscriptaddress,
 amsmath,amssymb,
 aps,
 pra,
floatfix,
]{revtex4-2}

\usepackage{graphicx}
\usepackage{dcolumn}
\usepackage{bm}
\usepackage{hyperref}
\usepackage[capitalise]{cleveref}
\usepackage{mathtools}
\usepackage{physics}
\usepackage{booktabs}
\usepackage{siunitx}
\usepackage{tikz}
\usepackage[T1]{fontenc}
\usepackage{pgfplots}
\pgfplotsset{compat=1.18}

\AtBeginDocument{\RenewCommandCopy\qty\SI}

\newcommand{\citeref}[1]{Ref.~\citep{#1}}

\usetikzlibrary{arrows.meta}
\usetikzlibrary{calc}
\usetikzlibrary{math}
\usetikzlibrary{matrix}
\usepgfplotslibrary{groupplots}
\usepgfplotslibrary{external}
\usepgfplotslibrary{statistics}

\begin{document}

\preprint{APS/123-QED}

\title{Driving Exchange Interaction in Spin Qubits with Quasi-Zero Pulses}

\author{Julian D. Teske}
 \altaffiliation{These authors contributed equally}
 \email{julian.teske@q-ctrl.com}
\author{Remy L. Delva}%
 \altaffiliation{These authors contributed equally}
 \email{julian.teske@q-ctrl.com}
\author{Shobhan Kulshreshtha}%
\author{Yuval Baum} %
\affiliation{%
 Q-CTRL, Los Angeles, CA USA \& Berlin, Germany
}%
\author{Florian Luthi}
\author{Fahd A. Mohiyaddin}
\author{Rostyslav Savytskyy}
\author{Thomas Watson}
\affiliation{%
 Intel Corporation, Technology Research Group, Hillsboro, OR 97124, USA
}%
\author{Pranav S. Mundada}%
\affiliation{%
 Q-CTRL, Los Angeles, CA USA \& Berlin, Germany
}%

\date{\today}

\begin{abstract}
The implementation of high-fidelity quantum gates for spin qubits requires accurate control of exchange interactions between electrons confined in quantum dots, but pulse distortions can limit this control accuracy.
Although linear-dynamical distortions can be compensated for by appropriately convolving the control signal, determining the necessary convolution requires detailed knowledge of the distortion's transfer function, and therefore the calibration of numerous parameters. Alternatively, control pulses can be designed to have a net-zero time integral canceling out linear-dynamical pulse distortions.
We generalize net-zero pulse designs to quasi-zero pulses allowing net-positive but reduced time integrals. Using these pulse designs, we systematically develop complete gate sets for exchange-only qubits, and study the resulting tradeoffs between pulse duration, fidelity, and the required number of tunable parameters, both in simulation and experiment. We benchmark the optimized gate pulses on Intel's Tunnel Falls six-dot device and show they achieve fidelities similar to those obtained with a full filtering approach, with identical pulse durations and fewer tuning parameters. This reduction in complexity opens the door to fast and easily automated calibration schemes compatible with large-scale commercial quantum devices.

\end{abstract}

\maketitle
\section{\label{sec:introduction}Introduction}

Semiconductor spin qubits are among the most promising candidates for the realization of a fault-tolerant quantum computer, since they combine short gate times and high gate fidelities with small feature sizes \cite{review_spin_qubits_2023}.
The promise of integration into industrial semiconductor fabrication processes is propelled by the recent success in the fabrication of high-quality qubit arrays in \(\qty{300}{\mm}\) \cite{intel_fab_probing, intel_fab_qubit_arrays} and \(\qty{200}{\mm}\) \cite{infineon_fab} fabrication lines for silicon-germanium-based systems, and of all-silicon devices \cite{imec_fab_SiMOS} in a \(\qty{300}{\mm}\) fabrication line. The recent deployment of cryogenic controllers for exchange-only qubits further facilitates the scaling of spin qubits for building a fault-tolerant quantum computer~\cite{membersofthehrlquantumteam2026digitallycontrolledsiliconquantum}.

The exchange interaction is the primary mechanism used to entangle the spins of confined electrons in quantum dots. Consequently, manipulation of the exchange coupling is a commonly used technique to realize single- and two-qubit quantum gates in most spin qubit implementations \cite{review_spin_qubits_2023}. 
This interaction can be driven with electric base-band pulses, which manipulate the electrostatic confinement potential and consequently the electrons' wave-function overlap. 
The exchange-only qubit is designed as a spin qubit platform that is exclusively controlled via such electrical signals driving the exchange interaction \cite{original_eo_DiVincenzo, review_three_el_qubits}.
Exchange-only qubits have been used to realize high-fidelity single-qubit~\cite{HRL_sigle_qubit_gates_2018} and two-qubit \cite{HRL_2qubit_gates_2023, intel_parallel_execution} gates, and have also been shown to be scalable into two-dimensional arrays \cite{HRL_2dim_spin_array, membersofthehrlquantumteam2026digitallycontrolledsiliconquantum}.

The exchange interaction energy has an approximate exponential dependence on the voltage and is, therefore, extremely sensitive to any distortion of the electric signals. Over the years, researchers have developed and employed a combination of strategies to combat signaling imperfections. These strategies include the application of finite and infinite impulse response filters on control signals or mitigating the distortion effects with pre-distortions \cite{predistortions_supercond, predistortions_semicond}. However, these methods necessitate the tuning of numerous parameters to achieve high fidelities \cite{intel_parallel_execution}. Generally, a high number of tunable parameters leads to long tuning times and less stable outcomes due to parameter drift over time.

Another strategy to combat the impact of distortions is to design robust pulses that are insensitive to particular distortions. The net-zero pulse Ansatz provides pulses with a net-zero time integral, where positive and negative pulse segments cancel each other out. This design provides a simple low-parameter construction to mitigate effects of RC time constants and bias-tee filtering by ensuring that residual offsets, long settling times, and low frequency distortions cancel out. In the field of superconducting qubits, it has been successfully utilized to yield high-fidelity gates \cite{NZ_literature_Rol, NZ_linerature_2}. 

In this work, we implement a highly scalable distortion-mitigation strategy on a six-dot implementation of Intel's Tunnel Falls device \cite{intel_fab_qubit_arrays} operated as an exchange-only qubit platform~\cite{intel_parallel_execution}. We generalize the traditional net-zero pulse framework into a class of robust "quasi-zero" pulses that feature a net-positive but significantly reduced time integral. Combining open-loop hardware testing with numeric simulations via the Boulder Opal software package\cite{boulder_opal}, we map out the fundamental trade-offs governing pulse duration, gate fidelity, and calibration complexity. Finally, using blind randomized benchmarking, we demonstrate that these optimized quasi-zero gates achieve operational fidelities and gate durations comparable to deeply optimized, multi-parameter pre-distortion filters, while fundamentally minimizing the number of parameters requiring active experimental calibration.

The remainder of this paper is organized as follows: \cref{sec:net-zero} details the formulation of net-zero and quasi-zero pulse shapes alongside experimental evidence of elimination of pulse distortion effects. \cref{sec:results} evaluates the systematic benchmarking and performance trade-offs of the resulting gate sets. Finally, \cref{sec:conclusion} presents our concluding remarks and outlook.

\section{\label{sec:net-zero}Net-Zero pulses}

The starting point of our investigation is the observation of pulse distortions in control signals. Under the assumption that these distortions are linear-dynamical in nature, they can be described by the system's response to a voltage signal with the shape of a step function.
In the investigated device, this response \(v_{\text{res}}\) was measured in preliminary experiments and fitted to the functional form:
\begin{align} \label{eq:step_response}
    v_{\text{res}}(t) = \alpha \left(1 - \sum_i c_i e^{-t/\tau_i} \right),
\end{align}
where \(\alpha\) denotes the amplitude of the step function in time \(t\). The fitting parameters are the magnitudes \(c_i\) (and the timescales \(\tau_i\)), which assume values between 0.03 and 0.1 (\(\qty{4}{\ns}\) and \(\qty{1}{\ms}\)). The timescale parameters \(\tau_i\) are essential for the effects of the distortions. Short timescales on the order of nanoseconds describe distortions that fully arise and decay within a single pulse, while long timescales of microseconds introduce memory effects depending on the previously executed pulses or quantum gates. 

\begin{figure}[t]
\centering
\tikzsetnextfilename{pulse_definitions}
\begin{tikzpicture}[x=0.2\linewidth,y=0.2\linewidth]
    \node[inner sep=0pt] (russell) at (2.7,.9)
        {\includegraphics[width=.3\linewidth]{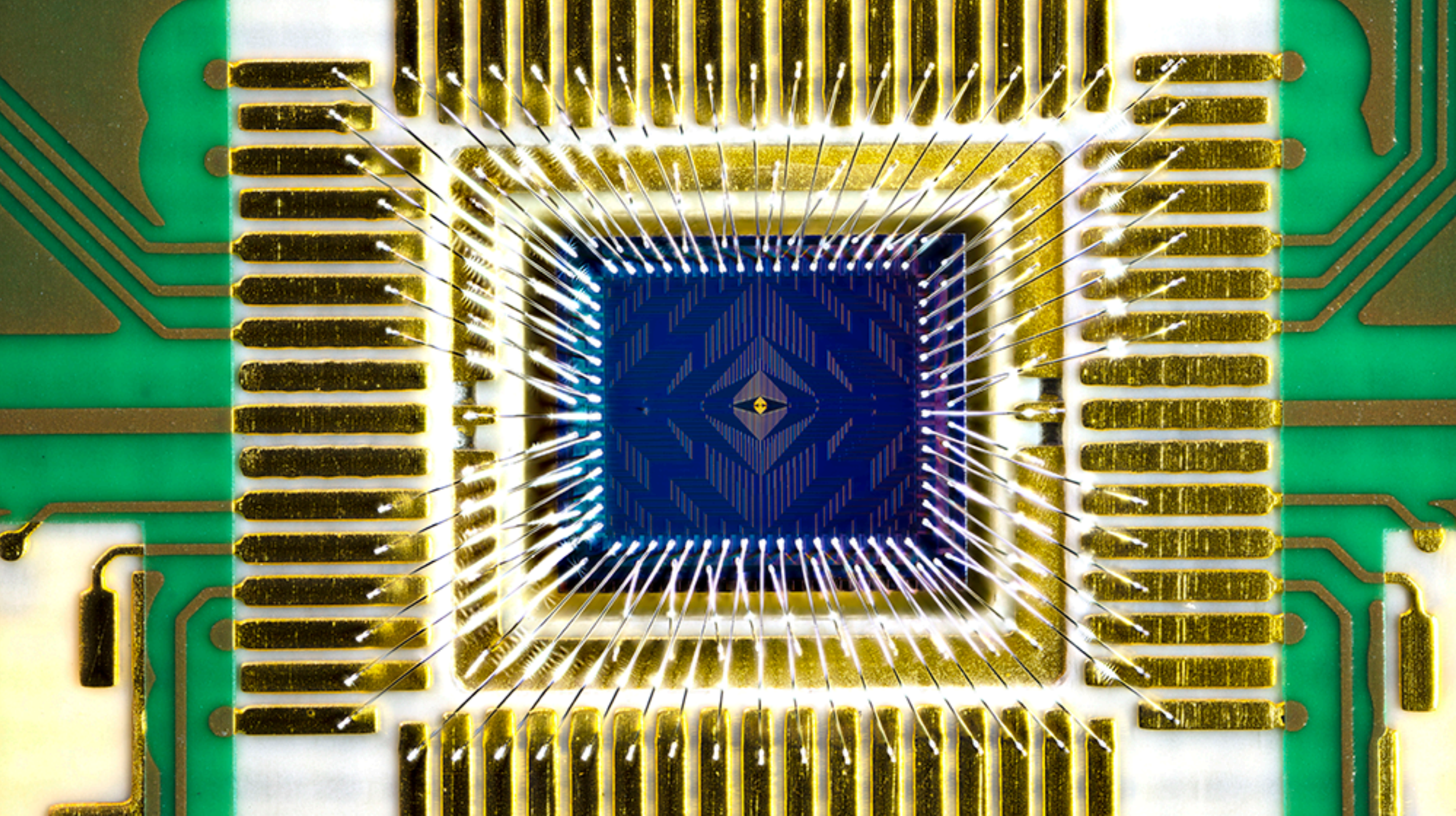}};
    \coordinate (d1) at (0,0);
    \coordinate (d2) at ($(d1) + (0.75,1)$);
    \coordinate (d3) at ($(d2) + (0.5,-1)$);
    \coordinate (d4) at ($(d3) + (1.25,-0.5)$);
    \coordinate (d5) at (4,0);
    \draw [thick,densely dotted,gray] (d1) -- (d1 -| d2) node [pos=0.5,below] {\(v=0\)} -- (d5);
    \draw [thick] (d1) |- (d2) |- (d3) |- (d4) |- (d5);
    \draw [thick,dotted] (d5) -- (d5 |- d2);
    \begin{scope}[Bar-Bar]
        \draw ($(d1) - (4pt,0)$) -- ($(d1 |- d2) - (4pt,0)$) node[pos=0.5,left] {\(A\)};
        \draw ($(d1 |- d2) + (0,4pt)$) -- ($(d2) + (0,4pt)$) node [pos=0.5,above] {\(t_{\text{pos}}\)};
        \draw ($(d2 |- d3) + (1pt,4pt)$) -- ($(d3) + (0,4pt)$) node [pos=0.5, above] {\(t_{\text{zero}}\)};
        \draw ($(d3 |- d4) + (0,-4pt)$) -- ($(d4) + (0,-4pt)$) node [pos=0.5,below] {\(t_{\text{neg}}\)};
        \draw ($(d3) + (-4pt,-1pt)$) -- ($(d3 |- d4) + (-4pt,0)$) node [pos=0.5,left] {\(\gamma A\)};
        \draw ($(d4 |- d5) + (0,4pt)$) -- ($(d5) + (-1pt,4pt)$) node [pos=0.5,above] {\(t_\text{space}\)};
    \end{scope}
    \begin{scope}[shift={(0,-0.75)},{Stealth[length=4]}-{Stealth[length=4]},gray]
        \draw (0,0.25) node [above] {\(v\)} -- (0,0) -- (0.25,0) node [pos=1,right] {\(t\)};
    \end{scope}
\end{tikzpicture}
\caption{Illustration of an atomic barrier pulse as a function of time. A rectangular exchange pulse would correspond to an amplitude ratio of \(\gamma=0\), while a net-zero pulse would correspond to \(\gamma=t_{\text{pos}}/t_{\text{neg}}\). Quasi-zero pulses correspond to other values of \(0 < \gamma < t_{\text{pos}}/t_{\text{neg}}\). In the following, we assume \(t_{\text{pos}}=t_{\text{neg}}\). Inset: picture of the Tunnel Falls device adopted from Ref.~\cite{intel_fab_qubit_arrays}.
}\label{fig:net_zero_pulses}
\end{figure}
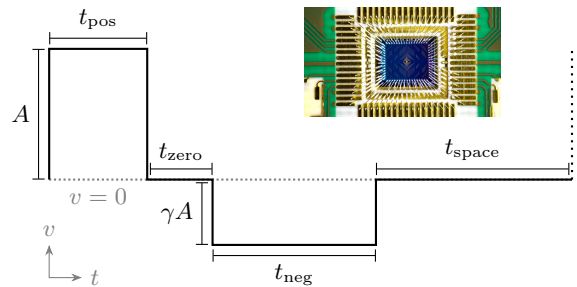

The considered exchange-only qubits are controlled with symmetric rectangular base-band pulses controlling barrier gates in pairs of coupled quantum dots \cite{symmetric_exchange_gate}. To control the time integral of our control pulses, we construct control signals by concatenating atomic pulses with positive and negative segments. The general form of these pulses is depicted in \cref{fig:net_zero_pulses}. Both the positive and negative segments can be scaled in duration and amplitude by changing the amplitude ratio \(\gamma\). After each segment, we include a temporal spacer to control for short-time distortions that could lead to bleed-through effects.

% figure: our net zero pulses

\begin{figure}[th]
    \centering
    \includegraphics[width=\linewidth]{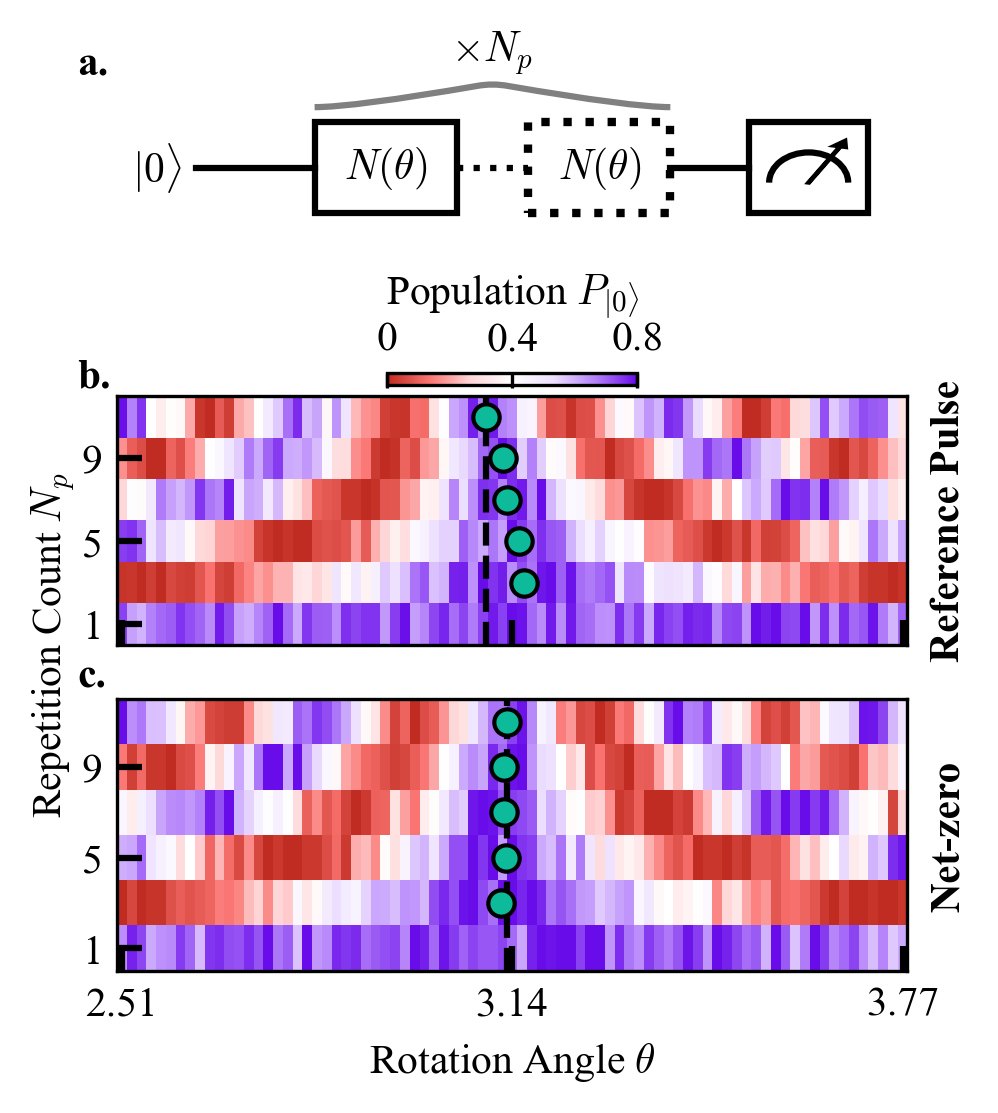}
    \caption{Pulse perturbation of net-zero and reference pulses. (a) Experiment description: The qubit is initialized, \(N_p\) repetitions of \(N(\psi)\) pulses are applied and the qubit is read out. Each \(N(\psi)\) pulse is tuned to rotate the qubit an angle \(\psi\) around the n-axis. (b, c) Measured fringe pattern of the reference pulse and the net-zero pulse. We plot the excitation probability as a function of the rotation angle \(\psi\) and number of pulse repetitions \( N_p\) which are incremented in multiples of 2. The teal circles mark the center of the fringe pattern for different repetition counts and black dashed lines are included as guide to the eye to spot the curvature of the fringes. }
    \label{fig:charge_acc}
\end{figure}

The evolution of an exchange-only qubit is described by the effective Hamiltonian
\begin{equation} \label{eq:eff_hamiltonian}
    H = J_z(v_z)\frac{\sigma_z}{2} + J_n(v_n)\frac{\sigma_n}{2},
\end{equation}
where \(J_i\) denotes the exchange interaction driven by voltage \(v_i\) for \(i \in \{z, n\}\). We considered \(v_i\) to be the voltage on a (virtual) barrier gate driving symmetric exchange gates. \(\sigma_z\) is the Pauli Z operator and \(\sigma_n\) generates a rotation around an axis that is tilted by \(\qty{120}{\degree}\) with respect to the Z-axis. The exponential relation between the exchange interaction and the barrier gate voltage is well-approximated over a large range of values by the function \(J_i(v_i) \propto e^{v_i/v_0}\). This relation strongly suppresses the exchange interaction at negative segments leading to an idling behavior and we therefore expect the negative pulse segments not to affect the evolution of the qubit's state. 

\begin{figure*}[t]
    \centering
    \includegraphics[width=\textwidth]{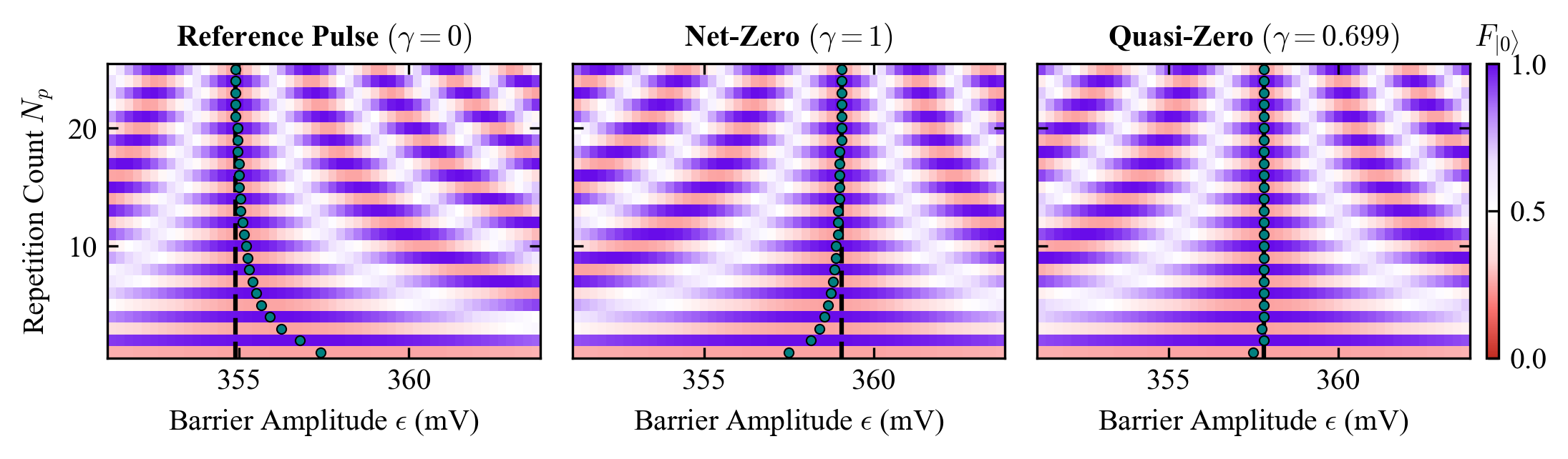}
    \caption{Comparison of fringe curvature induced by reference, net-zero and quasi-zero pulses. The plot displays the simulated ground-state fidelity $F_{\ket{0}}$ following the multi-pulse sequence outlined in \cref{fig:charge_acc} (a). The teal dots are the peaks along the central fringe and the black dashed line marks the calibrated amplitude to implement the \(N(\pi/2)\) gate. 
    We observe that the net-zero pulse bends the curvature of the central fringe and over-corrects the curvature on the considered time-scale. The optimal amplitude relation \(\gamma\) of a quasi-zero pulse is found by maximizing the alignment of the teal dots along the central fringe to a straight line. 
    }
    \label{fig:ar_tuning}
\end{figure*}

We experimentally verify the hypothesis that net-zero pulses can eliminate long-time pulse distortions by driving approximate \(\pi\) rotations around the N-axis of the exchange-only qubit as described in \cref{fig:charge_acc}~(a). The \(N(\pi)\) gate is realized with a single atomic pulse as plotted in \cref{fig:net_zero_pulses}. Thereby, variations of the amplitude near the calibrated value, for a \(\pi\) rotation, create Rabi fringes. For purely positive square pulses as a \textit{reference pulse} (\( \gamma=0 \)), we expect the long-time perturbations to cause systematic over-rotations at higher repetition numbers, whereas this effect should not occur for net-zero pulses. These time-dependent over-rotations are visible as a curvature in the Rabi fringes in \cref{fig:charge_acc}~(b). In agreement with our hypothesis, we observe that these systematic shifts of the fringe pattern vanish when employing net-zero pulses as shown in \cref{fig:charge_acc}~(c). This is a clear indication that the long-time distortions are partially caused by a linear-dynamical effect like charge accumulation, for example.

We study the influence of the new atomic pulse shape in simulations implemented with Boulder Opal \cite{boulder_opal}. The simulation is based on the effective Hamiltonian from \cref{eq:eff_hamiltonian} and uses a transfer function extracted from the step response in \cref{eq:step_response} with \(\alpha = 1\), \(c = (0.1,\,0.03,\,0.03,\,0.03)\) and \(\tau = (4,\,50,\,200,\,1000)\unit{\ns}\).
The simulations allow us to track the curvature of the central fringe more accurately and provide the insight that the use of net-zero atomic pulse shapes with \(\gamma=1\) leads to an over-correction of the perturbations shown in \cref{fig:ar_tuning}. We assume that this effect is caused by a bias due to the order of negative and positive segments, where the positive segments always precede the negative segments. We investigate a reduction of the negative amplitudes to avoid the over-correction by optimizing the alignment of the central fringe to a straight line. An ideal alignment would provide a pulse amplitude that does not show any time-dependent over or under-rotations. For typical configurations with the measured step response, an amplitude ratio around \(\gamma \approx 70\%\) seems to be ideal. For this amplitude ratio, the negative and positive curvatures cancel each other to the largest possible extent on the considered timescales.

\section{\label{sec:results}Quasi-Zero Gate Set Performance}

We realize the single-qubit gate set described in \citeref{HRL_sigle_qubit_gates_2018} by substituting our atomic pulse described in \cref{fig:net_zero_pulses} for the standard square pulse segment and optimizing the amplitude ratio \(\gamma\). We assess the quality of the resulting gate sets with blind randomized benchmarking \cite{HRL_sigle_qubit_gates_2018}. We also consider combinations of pre-distortions \cite{intel_parallel_execution} with our atomic pulses. In this case, we measure the performance after tuning all free pre-distortion parameters.

\begin{table*}[thbp]
    \centering
    \begin{tabular}{SSSSSSSSSSS}
        \toprule
        {Index} & {Pre-dist.} & {\# Filter param.} & {\(t_{\text{pos}}\ (\unit{\ns})\)} & {\(t_{\text{zero}}\ (\unit{\ns})\)} & {\(t_{\text{neg}}\ (\unit{\ns})\)} & {\(t_{\text{space}}\ (\unit{\ns})\)} & {\(t_{\text{total}}\ (\unit{\ns})\)} & {\(\gamma\)} & {Fidelity \((\unit{\percent})\)} & {Leakage \((\unit{\percent})\)} \\ \midrule
        1 & {None} & 0 & 10 & 20 & 0 & 0 & 30 & 0 & 99.66 & 0.01 \\
        2 & {Full} & 12 & 10 &20 & 0 & 0 & 30 & 0 & 99.96 & 0.01 \\ \hline
%        3 & \qty{30}{\ns} & 4 & 20 & 0 & 20 & 20 & 60 & 1 & 99.90 & 0.02 \\
%        4 & \qty{30}{\ns} & 4 & 10 & 20 & 10 & 20 & 60 & 1 & 99.85 & 0.03 \\ \hline
        3 & {None} & 0 & 10 & 0 & 10 & 10 & 30 & 0.25 & 99.93 & 0.01 \\
        4 & {None} & 0 & 10 & 0 & 10 & 20 & 40 & 0.5 & 99.94 & 0.01 \\
        5 & {None} & 0 & 20 & 0 & 20 & 20 & 60 & 0.4 & 99.91 & 0.02 \\
        % 5 & \qty{30}{\ns} & 4 & 20 & 20 & 0 & 0 & 40 & 0 & 99.93(1) & 0.01(1) \\
        6 & \qty{30}{\ns} & 4 & 10 & 0 & 10 & 20 & 40 & 0.525 & 99.90 & 0.01 \\
        7 & \qty{752}{\ns} & 2 & 10 & 0 & 10 & 20 & 40 & 0.55 & 99.95 & 0.01 \\
        8 & \qty{1}{\micro \second} & 2 & 10 & 0 & 10 & 20 & 40 & 0.55 & 99.95 & 0.01 \\
        \bottomrule
    \end{tabular}
    \caption{
    Experimental data collected on 6-dot Intel device: Single-qubit randomized benchmarking fidelity, leakage and segment durations of reference (\(\gamma=0\)), net-zero (\(\gamma=1\)) and quasi-zero (\(0<\gamma<1\)) pulses measured by blind randomized benchmarking. The amplitude ratios are optimized for the given parameter combinations on the experiment. }
    \label{tab:detailed_benchmarks}
\end{table*}

We test gate sets with different parameter combinations to systematically explore the properties of quasi-zero pulses and demonstrate the most important tradeoffs in the pulse design. The different parameters and the measured infidelities are summarized in \cref{tab:detailed_benchmarks}. We benchmark reference (\#1-2) and quasi-zero (\#3-8) pulses combined with a selection of pre-distortions and variations in the individual pulse segment durations. These segment durations can have a strong influence on the distortion dynamics, for example when long spacer times \(t_{\text{space}}\) partially cancel bleed-through between atomic pulses. The resulting total gate times affect the susceptibility to leakage caused by hyperfine noise. The different filter methods (\#6-8) cover filter choices selectively filtering pre-distortions on short and long timescales.

The detrimental effect of the pulse distortions is captured by using pulse setting \#1 as our reference pulse without any pre-distortions. Its fidelity is limited by the distortions to \(99.66\%\). This fidelity serves as a point of reference to measure whether the designed pulses reduce the pulse distortions. 
The target performance is established by the full filtering approach \cite{intel_parallel_execution} (setting \#2), where the fidelity is substantially increased up to \(99.96\%\) at the expense of introducing 12 tunable parameters. This pulse provides a reference point for the fidelity when pulse distortions are strongly suppressed.

In setting \#3, we consider a quasi-zero pulse that reaches a fidelity of \(99.93\%\) coming close to the performance of the fully filtered setting \#2 of the same total duration of \(\qty{30}{\ns}\), but without any required tuning parameters.
The fidelity can be further increased by increasing the spacing time after the negative segment to \qty{20}{\ns} bringing the total pulse duration to \qty{40}{\ns} while achieving a fidelity of \(99.94 \% \) as in setting \#4. Also increasing the driving segments to \qty{20}{\ns} as in setting \#5 reduced the fidelity to \(99.91\%\). This demonstrates the tradeoff between longer pulse durations and higher driving amplitudes. Long pulses are affected more strongly by magnetic noise, while the exponential relation between the exchange interaction and the driving voltage increases the susceptibility to charge noise at higher driving amplitudes~\cite{HRL_hyperfine_noise}.

We further explore the combination of quasi-zero pulses with pulse filters. Adding a filter on a short time scale of \qty{30}{\ns}, as in setting \#6, reduced the fidelity to \( 99.90\%\) indicating that the quasi-zero pulses are already sufficient to reduce distortions on short timescales.
In settings \#7 and \#8, we observe that a combination of net-zero pulses with long-time pre-distortions on the timescales of \qty{752}{\ns} or \qty{1000}{\ns} achieved a fidelity of \( 99.95\%\) coming close to the performance of the fully corrected pulse within measurement accuracy, while requiring only 2 tuning parameters. This high fidelity shows that the optimization of the amplitude ratio \(\gamma\) leads to pulses that do not fully cancel out errors on larger timescales. Overall, we observe that quasi-zero pulses can reach similar fidelities and durations as the best reference pulses, but with fewer parameters that need to be optimized.

\section{\label{sec:conclusion}Conclusion}

In this work, we demonstrated that quasi-zero control pulses provide a highly effective, low-complexity framework for mitigating linear-dynamical pulse distortions in exchange-coupled spin qubits. By systematically tuning the amplitude and duration of negative pulse segments, we bypassed the systematic overcompensation penalties inherent to strict net-zero designs while retaining their robust distortion-canceling benefits. This approach yields high-fidelity single-qubit gate sets validated via both Boulder Opal simulations and direct hardware implementation on Intel's Tunnel Falls processor.

With a systematic benchmarking of these gate sets, we mapped out the possible trade-offs between pulse duration, required tuning parameters and achieved gate fidelity. 
The optimal tradeoff is dictated by the interplay between hyperfine noise, which degrades performance at longer pulse durations, and charge noise, which is amplified at higher and therefore shorter pulse durations.
We also demonstrated the possibilities to combine pre-distortions with quasi-zero pulses. These combinations offer similar fidelities to the fully corrected pulse sets but with much fewer tunable parameters. The proposed quasi-zero pulses therefore offer a simple and robust solution to mitigate charge accumulation that can accelerate the tune up process for spin qubits compared to a fully tuned pre-distortions approach.

Further improvements of the presented gate-set performance could be based on a detailed analysis of the charge configuration at negative pulse values. This could include an investigation of the introduced deformation and displacement of the quantum dots and the influence on spurious coupling between other quantum dot pairs. We suggest an analysis of spurious crosstalk and how this affects the susceptibility to charge and hyperfine noise. The work can then be extended to two-qubit gates.

A comparison of pulse generators and fridge setups would provide insights regarding how strong, unwanted signal reflections and the conductor capacitance affect the pulse distortions compared to charge accumulation. Especially interesting is the comparison to cryo-electronics, where shorter transmission lines should lead to fewer pulse distortions while also the filtering capabilities will be severely restricted compared to an arbitrary waveform generator at room temperature making quasi-zero pulses even more attractive to mediate charge accumulation. 

The control-pulse simplifications demonstrated here pave the way for easily automated, large-scale calibration of exchange interactions in commercial quantum processors. We expect our findings to be transferable to other qubit types that rely on driving the exchange interaction to realize single qubit gates for the singlet-triplet qubit or two qubit gates for the Loss-DiVincenzo qubit.

\bibliography{apssamp}

\end{document}